\documentclass[onecolumn,10pt]{IEEEtran}
\usepackage{graphicx}
\usepackage{framed}
\usepackage{epstopdf}
\usepackage{amsmath}
\usepackage{amsthm}
\usepackage{amsfonts}
\usepackage[justification=centering]{caption}
\usepackage{soul} 
\usepackage{color, xcolor}
\theoremstyle{}
\newtheorem{theorem}{Theorem}
\newtheorem{lemma}{Lemma}
\newtheorem{definition}{Definition}

\newtheorem{example}{Example}

\usepackage{bm}
\usepackage{booktabs}
\usepackage{multirow}
\usepackage{algorithm}
\usepackage{algpseudocode}
\usepackage{subfigure}
\usepackage{amssymb}
\usepackage{cite}
\usepackage{multicol}
\usepackage{lipsum}
\usepackage{pstricks}
\usepackage[T1]{fontenc}
\usepackage{setspace}
\newcommand{\tabincell}[2]{\begin{tabular}{@{}#1@{}}#2\end{tabular}}
\begin{document}
\title{Improved Coded Caching Scheme for Multi-User Information Retrieval System}
\author{Junyi~Wang, Quan~Zang, Jinyu~Wang, Minquan~Cheng,~\IEEEmembership{Member,~IEEE,}

\thanks{J. Wang and Q. Zang are with the Key Laboratory of Cognitive Radio and Information Processing, Ministry of Education, Guilin University of Electronic Technology, Guilin 541004, China (e-mail:   wangjy@guet.edu.cn, quanzang@mails.guet.edu.cn). }
 
\thanks{J. Wang and M. Cheng are with the Key Laboratory of Education Blockchain and Intelligent Technology, Ministry of Education, and the Guangxi Key Laboratory of Multi-Source Information Mining and Security, Guangxi Normal University, Guilin 541004, China  (e-mail: mathwjy@163.com, chengqinshi@hotmail.com). }

}

\maketitle

\begin{abstract}
In this paper, we study the coded caching scheme for the $(L, K, M, N)$ multi-user information retrieval (MIR) system, which consists of a content library containing $N$ files, a base station (BS) with $L$ antennas that cannot access the library, and $K$ single-antenna users, each of which can cache at most $M$ files from the library. The users communicate with the others assisted by the BS to decode their required files. In this paper, we focus on designing a coded caching scheme with low communication latency measured by normalized delivery time (NDT), computational complexity, and subpacketizations. When $\frac{KM}{N}\geq L$ we first simply the precoding matrix in the downlink step to an identity matrix and use the multiple-antenna placement delivery array (MAPDA), which was originally proposed for the multiple-input single-output networks, to generate several new schemes for MIR system. Compared to the existing schemes, both the theoretical and numerical analyses show that our new schemes achieve much lower computational complexity and smaller subpacketizations with the same NDT. 
\end{abstract}


\section{Introduction}
\label{sect-introduction}
In recent years, the exponential growth of wireless traffic has put enormous pressure on wireless networks. This is due to the high temporal variability of network traffic, resulting in congestion during peak hours and underutilization during off-peak hours. Coded caching \cite{maddah2014fundamental} offers an effective way to alleviate network traffic by pre-populating content into each user's local cache during off-peak hours and applying coding theory to create more multicast opportunities during peak traffic hours. 

A caching system consists of two stages, namely the placement phase and the delivery phase. In the placement phase, the server places content in each user's cache without knowing their future demands. In the delivery phase, each user requests an arbitrary file, and the server broadcasts coded packets such that each user can decode its requested file based on its cached content.
The initial coded caching scheme was introduced by Maddah-Ali and Niesen(MN) in \cite{maddah2014fundamental} for a shared-link broadcast network, where a central server containing $N$ files of equal length is connected to $K$ users, each of which is capable of caching at most $M$ files through an error-free shared link. The objective of communication is to design a scheme that minimizes the transmission cost. The MN scheme achieves the order minimum communication load (i.e., the number of communication bits normalized by the file size) within a factor of 2 \cite{YMA}. In the case of uncoded placement, where each user directly stores a subset of the file bits, the MN scheme is precisely optimal when $N\geq K$ \cite{ERTCP,WTP}. Many studies have explored the coded caching problem across various network topologies, including Device-to-Device (D2D) networks \cite{JCM,JHJCM,ZXWL}, hierarchical networks \cite{ KNMD,THM}, combination networks \cite{JWTLCL,ZY,WJPT}, and arbitrary multi-server linear networks \cite{SMK}. In addition, coded caching has been extensively applied to wireless networks, including multiple-input single-output (MISO) networks \cite{EP, SCH, MB, SPSET, LBE, WCC, YWCC,EPDA,ASMST}, single-input-single-output (SISO) wireless networks \cite{CXHZW,NMA,HND,CAIC,FWNC,XTZ,Tao'TITic} and multiple-input multiple-output (MIMO) ntworks  \cite{MIMOcacheBC'19, TaoMIMOIC'17,MIMOIC'TIT23}.

In MISO coded caching problem, the authors in \cite{ASMST} first introduced the multi-user information retrieval (MIR) system \cite{NLLXA} which is a generic field of research exploring mechanisms to enable each user in the network to recover specific pieces of information that are either aggregated at a central master node \cite{XQK} or distributed across the network \cite{LMYA}. Specifically, a  $(L,K,M,N)$ MIR caching system includes a content library $N$ files, each of equal size, a BS equipped with $L$-antennas  that cannot access the library, and $K$ single-antenna users, each of which can cache at most $M$ files out of the $N$ files in the library. 
The users communicate with each other, assisted by the BS, to decode their required content. In a $(L,K,M,N)$ MIR coded caching scheme, the delivery phase consists of the uplink step and downlink step. That is, in the delivery phase, users transmit a portion of their accessible packets to the BS via a number of consecutive uplink transmissions in the uplink step; the BS then appropriately processes and combines the received signals, and forwards them back to the users in the downlink step. So the total communication load consists of both the uplink and downlink steps. Instead of the communication load, in this paper we use the concept of normalized delivery time (NDT) \cite{FWNC,XTZ,Tao'TITic} to characterize the efficiency of the communication. We prefer to design a scheme such that the NDT of the uplink/downlink step is as small as possible. 

In fact, the key points of designing a coded caching scheme for MIR system are constructing the precoding matrix used by each user in the uplink step and the precoding matrixt used by the BS in the downlink step at each transmission. By carefully designing these two classes of matrices, the authors in \cite{ASMST} first proposed a coded caching scheme for the MIR system, referred to as the ASMST scheme, where the NDT of the uplink/downlink step achieves the information-theoretic optimal in the MISO coded caching system under uncoded cache placement (i.e., each user directly stores some bits of files in the library)\cite{WTP,YMA} and one-shot linear delivery (i.e., each transmitted coded packet in a transmission could be decoded by any required user and its cache) \cite{LBE}. 

However, the ASMST scheme has a high computational complexity which is about $O((t+L) {t+L-1\choose t}^{3}{K\choose t+L} )$ in designing these precoding matrice and decoding the required files for users where $t=\frac{KM}{N}$. In addition, the other disadvantage of ASMST scheme is that, it usually needs to split each file into $F={K\choose t}{K-t-1\choose L-1}$ packets, where $F$ increases exponentially with the number of users $K$. This would become infeasible when $K$ is large. Therefore, designing a coded caching scheme for MIR system with smaller subpacketization will be a critical issue, especially for practical implementations.   
\subsection{Contributions and organization}
In this paper, we focus on designing  coded caching schemes for MIR system with lower computational complexities and lower subpacketizations. When $t=\frac{KM}{N}\geq L$, we find that under the same NDT as that of the ASMST scheme, the precoding matrix used by BS in the downlink step can be simplified to a identity matrix, i.e., the BS can directly transmit the received signals to the users through the downlink channel. This implies that we only need to study the precoding matrix used by each user in the uplink step. In addition, we  find that an interesting combinatorial structure called multi-antenna placement delivery array (MAPDA) introduced in \cite{YWCC}, which can be used to realize a coded caching scheme for MISO networks, can ensure that the precoding matrix in the uplink step always exists. So using the existing results of the MAPDAs in \cite{YWCC}, \cite{EPDA} , the following new schemes for the MIR system, denoted by Scheme 1, Scheme 2 and Scheme 3, can be obtained. Compared to the ASMST scheme, our new schemes have the following advantages.

$\bullet$ Our schemes achieve the same normalized delivery time (NDT) with smaller subpacketizations. Specifically, the NDTs of our schemes are $\tau_{\text{UL}}=\tau_{\text{DL}} =\frac{K(1-M/N)}{t+L}$, which exactly match those of the ASMST scheme. The subpacketization of the ASMST scheme grows exponentially with $K$, our schemes significantly reduce it, with subpacketization growing either subexponentially or linearly with $K$.

$\bullet$ Our schemes have much lower computational complexity when $K$ is sufficiently large. Specifically, it is approximately $O((t+L-1)^{3t-2}\cdot \alpha^{\frac{t+L}{\alpha}}\cdot K^{(t+L)\cdot\frac{\alpha-1}{\alpha}} )$ times larger than Scheme 1, $O((t+L-1)^{3t-2}\cdot m^{\frac{t}{m}}\cdot K^{L+t\cdot\frac{m-1}{m}} )$ times larger than Scheme 2, and $O((t+L-1)^{3t-2}\cdot K^{L+t-1})$ times larger than Scheme 3, where $\alpha=\gcd(K,t,L)$, $m\leq L$.

The rest of this paper is organized as follows. Section~\ref{sec-pre} describes the system model. Section \ref{main results} presents MAPDA for the MIR system and give the performance analyses. Some proofs can be found in Section \ref{proof}. Finally, we conclude our work in Section \ref{conclusion}.
 
 \subsection{Notations}
In this paper, the following notations will be used unless otherwise stated.

\noindent $\bullet$ $[a:b]:=\left\{ a,a+1,\ldots,b\right\}$ and $[a]:=\left\{ 1,2,\ldots,a\right\}$.  $|\cdot|$ denotes the cardinality of a set.

\noindent $\bullet$  We use the notation $a|q$ if $a$ is divisible by $q$ and $a\nmid q$ otherwise.  

\noindent $\bullet$ $gcd(a,b)$ denotes the greatest common divisor of $a$ and $b$.

\noindent $\bullet$  For any positive integers $n$ and $t$ with $t<n$, let ${[n]\choose t}=\{\mathcal{T}\ |\   \mathcal{T}\subseteq [n], |\mathcal{T}|=t\}$, i.e., ${[n]\choose t}$ is the collection of all distinct $t$-subsets of $[n]$.

\noindent $\bullet$ Given any $F\times m$ array $\mathbf{P}$, for any integers $i\in[F]$ and $j\in [m]$, $\mathbf{P}(i,j)$ represents the element located in the $i^{\text{th}}$ row and the $j^{\text{th}}$ column of $\mathbf{P}$; $\mathbf{P}(\mathcal{ V},\mathcal{T})$ represents the subarray generated by the row indices in $\mathcal{V}\subseteq [F]$ and the columns indices in $\mathcal{T}\subseteq [m]$. 

\noindent $\bullet$ We use $\mathbf{A}^{\mathrm{T}}$ and $\mathbf{A}^{*}$ to represents the transpose and conjugate-transpose (Hermitian) of matrix $\mathbf{A}$, respectively.
 
\section{Preliminaries}\label{sec-pre}
In this section, we first intrduce the MIR system model and the existing results. Then we present the concept of the multi-antenna placement delivery array and some existing results which will be useful in this paper.
\subsection{Model system}
\begin{figure}[http!]
\centering
\includegraphics[scale=0.5]{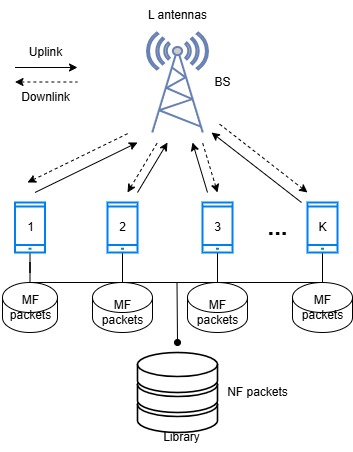}
\caption{Multi-user information retrieval (MIR) system.}
\label{figure-system}
\end{figure}

As depicted in Figure \ref{figure-system}, a $(L,K,M,N)$ MIR system consists of a content library containing $N$ files $\mathbf{W}_n$, $n\in [N]$, each of which has $B$ bits, a BS with $L$-antennas that cannot access the library, and $K$ single antenna users each of which can cache at most $M$ files of $N$ files in the library. The users communicate with each other, assisted by the BS, to decode their required files.
An $F$-division $(L,K,M,N)$ MIR coded caching scheme consists of the following phases.

$\bullet$ {\bf Placement phase:} Each file is divided into $F$ packets, i.e., $\mathbf{W}_{n}=(\mathbf{W}_{n,1},\mathbf{W}_{n,2},\ldots,\mathbf{W}_{n,F})$ and each packet $\mathbf{W}_{n,f}\in\mathbb{F}_{2^B}$ for $n\in [N], f\in[F]$. Each user caches at most $MF$ packets. Denote the cached contents at user $k$ as $\mathcal{Z}_{k}$. In this paper, we consider the uncoded placement where every user directly caches some packets. 

$\bullet$ {\bf Delivery phase:} Each user $k$ randomly demands a file $\mathbf{W}_{d_k}$ where $d_k\in[N]$ and $k\in[K]$.   Let $\mathbf{d}\triangleq (d_1,\ldots,d_K)$ denote the demand vector. In order to satisfy the users' demands, the communication consists of the the following two steps, i.e., the uplink step and downlink step. 
\begin{itemize}
\item[--]{\bf In the uplink step}, users transmit a portion of their accessible contents to the BS via a number of consecutive uplink transmissions. More precisely, we first encode each packet to a \emph{coded packet} denoted by  $\tilde{ \mathbf{W}}_{n,f}=\psi( \mathbf{W}_{n,f})\in\mathbb{C}^{\tilde{B}}$, where $\psi$ is the coding function for the Gaussian channel with rate 
\begin{equation}
\label{eq-rate-packet}
\frac{B}{\tilde{B}}=\log P+o(\log P),
\end{equation} e.g., random Gaussian coding. Here each coded packet contains $\tilde{B}$ complex symbols and carries one degree-of-freedom (DoF). The whole uplink communication contains $S_{\text{UL}}$ transmissions, each of which consists of $\tilde{B}$  complex  symbols (i.e.,  $\tilde{B}$ time slots). For each $s \in [ S_{\text{UL}}]$, user $k$ where $k\in[K]$ sends the linear combinations of the requested packets, i.e.,   
\begin{equation}\label{T-sig}
\mathbf{x}_k(s)=\sum_{j\in[r_s]}v_{k,j}^{(s)}\tilde{\mathbf{W}}_{n_j,f_j},
\end{equation}
where $v_{k,j}^{(s)}$ can be chosen any value from $\mathbb{C}$ if user $k$ caches the packet $\tilde{\mathbf{W}}^{(s)}_{n_j,f_{j}}$, otherwise $v_{k,j}^{(s)}=0$. Assume that user $k \in \left[ K \right]$ is connected to the BS through the channel $ \mathbf{h}^{(s)}_{k}=(h^{(s)}_{l,k})\in\mathbb{C}^{L\times1} $, which is i.i.d., where $l\in[L]$, and $s\in[S_{\text{UL}}]$. After the transmission $s$, the BS receives the signal
\begin{align}
\label{eq-up-link}
\mathbf{y}_{\text{BS}}(s)& = \sum\limits_{k \in [K]} {{\mathbf{h} _k}\mathbf{x}_k(s)}  + \mbox{\boldmath$\epsilon$}_{\text{BS}}(s)
=(\mathbf{h}^{(s)}_{1}\ \mathbf{h}^{(s)}_{2}\ \cdots\ \mathbf{h}^{(s)}_{K})\left(
\begin{array}{c}
\mathbf{x}_1(s)\\
\vdots\\
\mathbf{x}_K(s)
\end{array}\right)+ \mbox{\boldmath$\epsilon$}_{\text{BS}}(s)\nonumber\\
&:=
\left(
\begin{array}{cccc}
h_{1,1}^{(s)}&h_{1,2}^{(s)}&\cdots&h_{1,K}^{(s)}\\
\vdots&\vdots&\ddots&\vdots\\
h_{L,1}^{(s)}&h_{L,2}^{(s)}&\cdots &h_{L,K}^{(s)}
\end{array}\right)
\left(
\begin{array}{cccc}
v_{1,1}^{(s)}&v_{1,2}^{(s)}&\cdots&v_{1,r_s}^{(s)}\\
\vdots&\vdots&\ddots&\vdots\\
v_{K,1}^{(s)}&v_{K,2}^{(s)}&\cdots &v_{K,r_s}^{(s)}
\end{array}\right)
\left(
\begin{array}{c}
\tilde{\mathbf{W}}^{(s)}_{n_1,f_{1}}\\
\vdots\\
\tilde{\mathbf{W}}^{(s)}_{n_{r_s},f_{r_s}}
\end{array}\right)
+\mbox{\boldmath$\epsilon$}_{\text{BS}}(s)\nonumber\\
&:=\mathbf{H}^{(s)}\mathbf{V}^{(s)}\left(
\begin{array}{c}
	\tilde{\mathbf{W}}^{(s)}_{n_1,f_{1}}\\
	\vdots\\
	\tilde{\mathbf{W}}^{(s)}_{n_{r_s},f_{r_s}}
\end{array}\right)
+\mbox{\boldmath$\epsilon$}_{\text{BS}}(s):=\mathbf{R}^{(s)}
\left(
\begin{array}{c}
\tilde{\mathbf{W}}^{(s)}_{n_1,f_{1}}\\
\vdots\\
\tilde{\mathbf{W}}^{(s)}_{n_{r_s},f_{r_s}}
\end{array}\right)
+\mbox{\boldmath$\epsilon$}_{\text{BS}}(s).
\end{align} By assuming  $P$ is sufficiently large, the noise can be ignored. Thus $\mathbf{y}_{\text{BS}}(s)$ is denoted by
\begin{align*}
\mathbf{y}'_{\text{BS}}(s) =
\mathbf{R}^{(s)}
\left(
\begin{array}{c}
\tilde{\mathbf{W}}^{(s)}_{n_1,f_{1}}\\
\vdots\\
\tilde{\mathbf{W}}^{(s)}_{n_{r_s},f_{r_s}}
\end{array}\right).
\end{align*}
To evaluate the transmission efficiency of the scheme, we adopt the same metric named \emph{normalized delivery time} (NDT) as in \cite{FWNC,XTZ,Tao'TITic}, which is defined as
\begin{align}\label{eq-original-NDT}
\tau(M )\triangleq
\lim_{P\to\infty}\lim_{V\to\infty}\sup\frac{\max_{{\bf d}\in[N]^K}T}{V/\log P},
\end{align} where $T$ is the total transmissions in the whole uplink communication. Since each file contains $F$ packets, each of file has $B$ bits and there are a total of $S_{\text{UL}}\tilde{B}$ time slots, by \eqref{eq-rate-packet} the equation \eqref{eq-original-NDT} can be written as 
\begin{align}\label{eq-comput-NDT}
\tau_{\text{UL}}(M )=\lim_{P\to\infty}\lim_{V\to\infty}\frac{S_{\text{UL}}\tilde{B}}{BF/\log P}
=\lim_{P\to\infty}\frac{S_{\text{UL}}}{F}\cdot\frac{\log P}{\log P+o(\log P)}
=\frac{S_{\text{UL}}}{F}.
\end{align}
From \eqref{eq-comput-NDT}, NDT can represent the maximal normalized number of transmitted files over all possible demands in the interference channel and the high signal-to-noise ratio (SNR) regime.  

\item[--]{\bf In the downlink step}, Assuming that the BS possesses sufficient computation capability to process and enough memory to store all received signals. Then the BS appropriately processes and combines the received signals, and forwards them back to the users. Assume that the downlink communication contains $S_{\text{DL}} $ transmissions. After receiving all the signals $\mathbf{y}'_{\text{BS}}(s)$ where $s\in [S_{\text{UL}}]$, the BS can send the signal $\mathbf{x}_{\text{BS}}(s')$ generated by its received signals to users, ensuring that each user can decode their requested file based on their cached content and the received signals. Specifically, for each $s'\in [S_{\text{DL}}]$, $K$ users receive the following signals
\begin{align}
\label{eq-down-link}
\mathbf{Y}(s')=\mathbf{y}'_{[K]}(s')&=
\left(
\begin{array}{c}
\mathbf{y}'_{1}(s')\\
\vdots\\
\mathbf{y}'_{K}(s')
\end{array}
\right)=\mathbf{M}^{(s')}\mathbf{x}_{\text{BS}}(s')=\mathbf{M}^{(s')}\mathbf{U}^{(s')}
\left(
\begin{array}{c}
\mathbf{y}'_{\text{BS}}(1)\\
\vdots\\
\mathbf{y}'_{\text{BS}}(S_{\text{UL}})
\end{array}
\right)+\mbox{\boldmath$\epsilon$}_{[K]}(s')\nonumber\\
&:=\left(
\begin{array}{cccc}
m_{1,1}^{(s')}&m_{1,2}^{(s')}&\cdots&m_{1,L}^{(s')}\\
\vdots&\vdots&\ddots&\vdots\\
m_{K,1}^{(s')}&m_{K,2}^{(s')}&\cdots &m_{K,L}^{(s')}
\end{array}\right)
\left(
\begin{array}{cccc}
u_{1,1}^{(s')}&u_{1,2}^{(s')}&\cdots&u_{1,LS_{\text{UL}}}^{(s')}\\
\vdots&\vdots&\ddots&\vdots\\
u_{L,1}^{(s')}&u_{L,2}^{(s')}&\cdots &u_{L,LS_{\text{UL}}}^{(s')}
\end{array}\right)
\left(\begin{array}{c}
\mathbf{y}'_{\text{BS}}(1)\\
\vdots\\
\mathbf{y}'_{\text{BS}}(S_{\text{UL}})
\end{array}\right)
+\mbox{\boldmath$\epsilon$}_{[K]}(s')
\nonumber\\
&:=\left(\begin{array}{cccc}
b_{1,1}^{(s')}&b_{1,2}^{(s')}&\cdots&b_{1,LS_{\text{UL}}}^{(s')}\\
\vdots&\vdots&\ddots&\vdots\\
b_{L,1}^{(s')}&b_{L,2}^{(s')}&\cdots &b_{L,LS_{\text{UL}}}^{(s')}
\end{array}
\right)
\left(
\begin{array}{c}
\mathbf{y}'_{\text{BS}}(1)\\
\vdots\\
\mathbf{y}'_{\text{BS}}(S_{\text{UL}})
\end{array}
\right)
+\mbox{\boldmath$\epsilon$}_{[K]}(s')\nonumber\\
&:=\mathbf{B}^{(s')}
\left(
\begin{array}{c}
\mathbf{y}'_{\text{BS}}(1)\\
\vdots\\
\mathbf{y}'_{\text{BS}}(S_{\text{UL}})
\end{array}
\right)
+\mbox{\boldmath$\epsilon$}_{[K]}(s').
\end{align}
Similarly to the process of uplink, by our assumption that $P$ is large enough each user $k\in[K]$ can decode the signal $\psi^{-1}(\mathbf{y}'_{k}(s'))$  with an error probability exponentially decreasing to zero with the NDT defined as follows.Since each file contains $F$ packets, each of which has $B$ bits and there are a total of $S_{\text{DL}}\tilde{B}$ time slots, \eqref{eq-original-NDT} can be written as
\begin{align}\label{eq-comput-NDT-DL}
\tau_{\text{DL}}(M )=\lim_{P\to\infty}\lim_{V\to\infty}\frac{S_{\text{DL}}\tilde{B}}{BF/\log P}
=\lim_{P\to\infty}\frac{S_{\text{DL}}}{F}\cdot\frac{\log P}{\log P+o(\log P)}
=\frac{S_{\text{DL}}}{F}.
\end{align}

\end{itemize}

From the above introduction, for simplicity we omit the coding process  $\psi$  and the noises in the uplink and the downlink steps in this paper. Clearly we prefer to design a scheme with the optimal NDT for up load defined as 
\begin{align*}
\tau^{*}_{\text{UL}}(M )\triangleq\inf\{ \tau_{\text{UL}}(M )|\tau_{\text{UL}}(M )\ \text{is achievable} \} .
\end{align*} and the optimal NDT for down load defined as
\begin{align*}
\tau^{*}_{\text{DL}}(M )\triangleq\inf\{\tau_{\text{DL}}(M )|\tau_{\text{DL}}(M )\ \text{is achievable} \}. 
\end{align*} 

\begin{lemma}[\cite{ASMST}]
\label{lemma-1}
For any positive integers $L$, $K$, $M$ and $N$ satisfying that $t=\frac{KM}{N}$ is an integer and $t+L\leq K$, there exists an $(L,K,M,N)$ MIR coded caching scheme with subpacketization $F={K\choose t}{K-t-1\choose L-1}$,  the NDTs $\tau_{\text{UL}}= \tau_{\text{DL}}=\frac{{K - t}}{{t + L}}$.
\end{lemma}

Let $t=KM/N$. Now let us introduce the sketch of the ASMST scheme in \cite{ASMST} as follows.

$\bullet$ {\bf Placement phase:} Each file is divided into $F={K\choose t}{K-t-1\choose L-1}$ packets, i.e., $\mathbf{W}_{n}=(\mathbf{W}_{n,\mathcal{T},\mathcal{L}})_{\mathcal{T}\in{[K]\choose t},\mathcal{L}\in{[K-t-1]\choose L-1}}$. Each user $k$ caches $\mathcal{Z}_{k}=\{\mathbf{W}_{n,\mathcal{T},\mathcal{L}}\ |\ k\in\mathcal{T}, \mathcal{T}\in{[K]\choose t},\mathcal{L}\in{[K-t-1]\choose L-1}\}$.

$\bullet$ {\bf Delivery phase:} We assume the demand vector ${\bf d}=(d_1,d_2,\ldots,d_K)$. For each $(t+L)$-subset of $[K]$, denoted by $\mathcal{U}$, each user in $\mathcal{U}$ tansmits ${t+L-1\choose t}$ times to the BS. At each transmission, $t+L$ users send signals to the BS where each of $t+1$ users just sends a signal packet required by other users from $\mathcal{U}$ and each of the left $L-1$ users sends a linear combination of $t+1$ packets required by other users from $\mathcal{U}$. So in each transmission,  the computational complexity is about $O\left( (t+1)(L-1)\right)$, and the total computational complexity for the subset $[t+L]$ is 
\begin{align}
\lambda_1=O\left( {t+L-1\choose t}(t+1)(L-1)\right).
\end{align}
By the above uplink transmission, the BS receives exactly ${t+L-1\choose t}L$ coded signals with the $L$ antennas. The authors used the same downlink transmission strategy as in \cite{XTZ}, i.e., they must first generate all possible coded signals, each of which has $(t+1)$ unique required packets, and then use these to create the desired coded signals to users for $L$ antennas, each of which has ${t+L\choose t+1}(t+1)$ unique required packets.
In these two steps, we can compute that the computational complexity are  
\begin{align}
\lambda_2=O\left({t+L\choose t+1} L\right),\ \ \ \ \  \lambda_3=O\left(2L{t+L\choose t+1}\right),
\end{align}respectively. Finally according to the after mentioned downlink transmission strategy, each user can decode its ${t+L-1\choose t}$  required packets with the computational complexity $O({t+L-1\choose t}^{3})$. So the total computational complexity of $t+L$ users is  
\begin{align}
\lambda_4=O\left((t+L){t+L-1\choose t}^{3}\right). 
\end{align}  
When $\mathcal{U}$ gets all the $(t+L)$-subsets of $[K]$, the total computational complexity of this scheme is 
\begin{align}
	\begin{split}
		\lambda_{\text{ASMST}}=&(\lambda_1+\lambda_2+\lambda_3+\lambda_4){K\choose t+L}\\
		=&\left( O\left( {t+L-1\choose t}(t+1)(L-1)\right) + O\left( {t+L\choose t+1}L \right) \right. \\  
		& \left. + O\left(2L{t+L\choose t+1}\right) + O\left((t+L) {t+L-1\choose t}^{3} \right) \right) {K\choose t+L}\\
		&\approx O\left((t+L) {t+L-1\choose t}^{3}{K\choose t+L} \right). 
	\end{split}
\end{align}

\subsection{Multi-antenna Placement Delivery Array}
In this subsection, we will introduce a useful concept called multiple-antenna placement delivery array (MAPDA)  proposed in \cite{YWCC} which can be used to characterize the placement strategy and delivery strategy for the MISO caching system.
\begin{definition}[\cite{YWCC}]\rm
\label{def-MAPDA}
For any positive integers $L$, $K$, $F$, $Z$ and $S$, an $F\times K$ array $\mathbf{P}$ that is composed of $``*"$ and $[S]$ is called $(L,K,F,Z,S)$ multiple-antenna placement delivery array (MAPDA) if it satisfies the following conditions

\noindent C$1$. The symbol $``*"$ appears $Z$ times in each column;

\noindent C$2$. Each integer occurs at least once in the array;

\noindent C$3$. Each integer $s$ appears at most once in each column;

\noindent C$4$. For any integer $s\in[S]$, define  $\mathbf{P}^{(s)} $
to be the subarray of $\mathbf{P}$ including the rows and columns containing $s$, and let $r'_s\times r_s$ denote the dimensions of $\mathbf{P}^{(s)}$.  The number of integer entries in each row  of $\mathbf{P}^{(s)}$ is less than or equal to $L$, i.e.,
\begin{align}\label{C4}
\left|\{k_1\in [r_s]  |\ \mathbf{P}^{(s)}(f_1,k_1)\in[S]\}\right|\leq L, \ \forall f_1 \in [r'_s].
\end{align}\hfill $\square$
\end{definition}
\begin{example}\label{ex-1}
When $K= 6$, $F=3$ and $Z=1$, the following $3\times 6$ array $\mathbf{P}$ satisfies conditions of Definition \ref{def-MAPDA}. For instance when $s=1$, we have the following $\mathbf{P}^{(1)}$ which satisfies condition C4. So $\mathbf{P}$ is a $(2,6,3,1,3)$ MAPDA. 		\hfill $\square$ 
\begin{align*}
	\mathbf{P}=\left(\begin{array}{cccccc}
		* & 1 & 2 & * & 1 & 2 \\
		1 & * & 3 & 1 & * & 3 \\
		2 & 3 & * & 2 & 3 & * 
	\end{array}\right), \ \ \ 	
	\mathbf{P}^{(1)}=\left(\begin{array}{cccc}
		* & 1 & * & 1 \\
		1 & * & 1 & * \\
	\end{array}\right).
\end{align*}  
\end{example}  
There are many construction of MAPDA in \cite{MB,SPSET,YWCC,EPDA}. Here we list some useful results of MAPDAs which have smaller subpacketizations for some parameters $K$, $t$ and $L$. 
\begin{lemma}[Existing MAPDAs]\rm
\label{le-existing}
For any positive integers $K$, $t$ and $L$ with $t+L\leq K$, there exist an $(L,K,F,Z,S)$ MAPDA with the following parameters:
\begin{itemize}\label{lemma2-MAPDAs}
\item When $t+L\leq K$, $F=\frac{t+L}{\alpha} \binom{K/\alpha}{(t+L)/\alpha}$, $Z=\frac{t}{\alpha}\binom{K/\alpha-1}{(t+L)/\alpha-1}$ and $S=
\frac{K-t}{\alpha}{K/\alpha\choose (t+L)/\alpha}$ where $\alpha=\gcd(K,t,L)$ \cite{EPDA};
\item When $t+L<K$ and $m\leq L$, $F=\beta {{K/m\choose t/m}}$, $Z=\frac{\beta t}{K} {{K/m\choose t/m}}$ and $S= \text{sgn}(\frac{t}{m}+1,\frac{m}{L})\cdot$ $l\cdot\frac{K-t}{t+m} {K/m\choose t/m}$ where  $l=\frac{m}{\gcd(m,L-m)}$, $\beta =(\text{sgn}(\frac{t}{m}+1,\frac{m}{L})+\frac{L-m}{m})l$, $m|K$ and $m|t$ \cite{YWCC};
\item When  $L=K-t$, $F=K$, $Z=t$ and $S=K-t$ \cite{YWCC}.
\end{itemize}
\end{lemma}

\section{Main results}\label{main results}
\label{sect-main}
In this section, we first present our main result, i.e., a coded caching scheme for the MIR system. Subsequently, we demonstrate our advantages, which include smaller subpacketization and lower computational complexity compared to existing schemes.

From \eqref{eq-up-link} and \eqref{eq-down-link}, the key point of designing a scheme for MIR system is how to design the precoding matrices for both the uplink and downlink steps (i.e., $\mathbf{V}^{(s)}$ and $\mathbf{U}^{(s')}$). By linear algebra, we find out that when $t=\frac{KM}{N}\geq L$, the precoding matrix $\mathbf{U}^{(s')}$ can be simplified into an identity matrix. In this case, the BS directly transmits the received signals to the users through the downlink channel, meaning that we only need to consider the precoding matrix $\mathbf{V}^{(s)}$. To this end, we show that the MAPDA introduced in \cite{YWCC} can be used to generate the precoding matrix $\mathbf{V}^{(s)}$ to obtain the following scheme.
Compared to the ASMST scheme \cite{ASMST}, where the BS must first generate all possible coded signals, each of which has $(t+1)$ unique required packets, and then use these to create the desired coded signals to users for $L$ antennas, each of which has ${t+L\choose t+1}(t+1)$ unique required packets, our approach significantly reduces computational complexity without sacrificing the Sum-DoF.

\begin{theorem}
\label{th-1} 
Given a $(L, K, F, Z, S)$ MAPDA with $KZ\geq LF$,  there exists an $F$-division $(L, K, M, N)$ MIR coded caching scheme with memory ratio $\frac{M}{N}=\frac{Z}{F}$ and the optimal $\tau_{\text{UL}}=\tau_{\text{DL}}=\frac{S}{F}$.
\end{theorem}
By Theorem \ref{th-1} and Lemma \ref{lemma2-MAPDAs}, we have the folowing schemes for the MIR system.
\begin{theorem} 
\label{th-2}
For any positive integers $K$, $t$ and $L$ with $t+L\leq K$ and $L\leq t$, there exist $F$-division $(L,K,M,N)$ MIR coded caching schemes listed in Table \ref{tab-Theorem-2} with NDT $\tau_{\text{new}}=\frac{K(1-M/N)}{t+L}$.
\end{theorem}

{\begin{table*}[http!]
\caption{The new schemes with the  sum-DoF $t+L$ in Theorem \ref{th-2} where $m\in \mathbb{Z}^{+}$ and $m|K$.}
\centering
\label{tab-Theorem-2}
\begin{tabular}{|c|c|c|c|}
\hline
New schemes&Subpacketization & Original MAPDA&Parameter limitations\\
\hline
Scheme 1 & $\beta {{K/m\choose t/m}}$&\cite{YWCC} & \tabincell{c}{$t+L<K$, $m\leq L$,  $l=\frac{m}{\gcd(m,L-m)}$,\\[0.1cm] $\beta =(\text{sgn}(\frac{t}{m}+1,\frac{m}{L})+\frac{L-m}{m})l$}\\ \hline

Scheme 2 & $\frac{t+L}{\alpha} \binom{K/\alpha}{(t+L)/\alpha}$&\cite{EPDA} & \tabincell{c}{ $t+L\leq K$, $\alpha=\gcd(K,t,L)$}\\ \hline

Scheme 3&$K$&\cite{YWCC} & $L=K-t$ \\ \hline
\multicolumn{4}{l}{\small * $\text{sgn}(x,y)$ equals  $1$ if $ y=1$, and $x$ otherwise. }\\
\end{tabular}
\end{table*}}
When $L>t$, using Theorem \ref{th-2} we can silence $L-t$ antennas to obtain the following result.
\begin{theorem} 
\label{th-3}
For any positive integers $K$, $L$ and $t$ with $t<L$, there exist $F$-division $(L, K, M, N)$ MIR coded caching schemes listed in Table \ref{tab-Theorem-3} with memory size $M=\frac{ZN}{F}$, the sum-DoF $2t$ and the NDT $\tau_{\text{UL}}=\tau_{\text{DL}}=\frac{K-t}{2t}$.
\end{theorem}{\begin{table*}[http!]
	\caption{The new schemes with the sum-DoF $2t$ in Theorem \ref{th-3} where $m\in \mathbb{Z}^{+}$ and $m|K$.}
	\centering
	\label{tab-Theorem-3}
	\begin{tabular}{|c|c|c|c|}
		\hline
		New schemes&Subpacketization & Original MAPDA&Parameter limitations\\
		\hline
		Scheme 1 & $\beta {{K/m\choose t/m}}$&\cite{YWCC} & \tabincell{c}{$t<\frac{K}{2}$, $m\leq t$,  $l=\frac{m}{\gcd(m,t-m)}$,\\[0.1cm] $\beta =(\text{sgn}(\frac{t}{m}+1,\frac{m}{t})+\frac{t-m}{m})l$}\\ \hline
		
		Scheme 2 & $\frac{2t}{\alpha} \binom{K/\alpha}{2t/\alpha}$&\cite{EPDA} & \tabincell{c}{ $t\leq \frac{K}{2}$, $\alpha=\gcd(K,t)$}\\ \hline
		
		Scheme 3&$K$&\cite{YWCC} & $t=\frac{K}{2}$ \\ \hline
		\multicolumn{4}{l}{\small * $\text{sgn}(x,y)$ equals  $1$ if $ y=1$, and $x$ otherwise. }\\
	\end{tabular}
	\end{table*}}

\subsection{Performance Analyses}

Compared to the ASMST scheme in \cite{ASMST}, our schemes in Table \ref{tab-Theorem-2} have smaller subpacketization and lower computational complexity, while achieving the same NDT.

By Theorem \ref{th-2}, the NDTs of our schemes are $\tau_{\text{UL}}=\tau_{\text{DL}} =\frac{K(1-M/N)}{t+L}$, which exactly matche the NDTs of the ASMST scheme in  Lemma \ref{lemma-1}. Now let us consider the subpacketization. By Lemma \ref{lemma-1} we have the subpacketization of the $(L,K,M,N)$ ASMST scheme  $F_{\text{ASMST}}=\binom{K}{t}\binom{K-t-1}{L-1}$. By Table \ref{tab-Theorem-2} for the same parameters $L$, $K$, $M$ and $N$, Scheme 1, Scheme 2 and Scheme 3 have the subpacketizations $F_1=\beta\binom{K/m}{t/m}$, $F_2=\frac{t+L}{\alpha} \binom{K/\alpha}{(t+L)/\alpha}$ and $F_3=K$. Thus we observe that subpacketization of the ASMST scheme grows exponentially with the number of users $K$,  however the subpacketizations of the Scheme 1 and Scheme 2 grow  subexponentially with the number of users $K$, the subpacketization of the Scheme 3 grows linearly with the number of users $K$. Then we have the ratio of $F_{\text{ASMST}}$ and $F_1$ as follows. 
\begin{equation} \label{F-ratio-1}
	\begin{split}
		\frac{F_1}{F_{\text{ASMST}}}=\frac{\beta\binom{K/m}{t/m}}{\binom{K}{t}\binom{K-t-1}{L-1}}
		&\approx\frac{2^{\frac{K}{m}H(\frac{t}{K})}}{2^{KH(\frac{t}{K})}2^{(K-t-1)H(\frac{L-1}{K-t-1})}}\ \ \ \ (K\rightarrow \infty)\\
		&=2^{\frac{(1-m)}{m}KH(\frac{t}{K})-(K-t-1)H(\frac{L-1}{K-t-1})}\\
		&<2^{\frac{(1-m)}{m}KH(\frac{t}{K})}\\
		&\approx 2^{\frac{(1-m)}{m}K}.\ \ \ \ \ (\text{Fixed memory ratio } M/N)
	\end{split}
\end{equation} From \eqref{F-ratio-1} we can see that Scheme 1 has an exponential reduction in subpacketization when $K$ tends to infinity compared to the ASMST scheme, the subpacketization of Scheme 1 is about  $2^{\frac{(1-m)}{m}K}$ times smaller than that of the ASMST scheme.

Similarly, the ratio of the subpackektizations $F_{\text{ASMST}}$ and $F_1$ is
\begin{equation} \label{F-ratio-2}
	\begin{split}
		\frac{F_2}{F_{\text{ASMST}}}=\frac{\frac{t+L}{\alpha} \binom{K/\alpha}{(t+L)/\alpha}}{\binom{K}{t}\binom{K-t-1}{L-1}}
		&\approx
		\frac{2^{\frac{K}{\alpha}H(\frac{t+L}{K})}}{2^{KH(\frac{t}{K})}2^{(K-t-1)H(\frac{L-1}{K-t-1})}}\ \ \ \ (K\rightarrow \infty)\\
		&= 2^{\frac{K}{\alpha}H(\frac{t+L}{K})-KH(\frac{t}{K})-(K-t-1)H(\frac{L-1}{K-t-1})}\\
		&<2^{\frac{K}{\alpha}H(\frac{t+L}{K})-KH(\frac{t}{K})}\\
		&\approx2^{K\cdot (\frac{1-\alpha}{\alpha})}.\ \ \ \ \ (\text{Fixed memory ratio } M/N\ \text{and}\ L)
	\end{split}
\end{equation}From \eqref{F-ratio-2} we can see that Scheme 2 has an exponential reduction in subpacketization when $K$ tends to infinity compared to the ASMST scheme,  the subpacketization of Scheme 2 is about  $2^{K\cdot (\frac{1-\alpha}{\alpha})}$ times smaller than that of the ASMST scheme.

\begin{figure}[http!]
	\centering
	\includegraphics[scale=0.4]{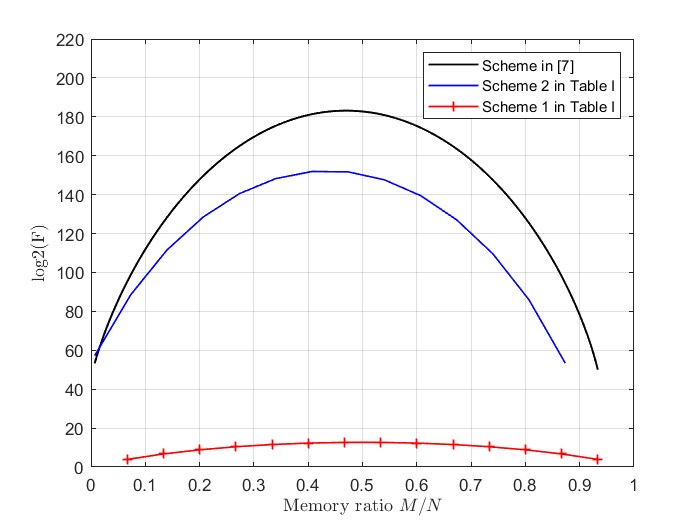}
	\caption{Subpacketization comparison between the schmes in Table \ref{tab-Theorem-2} and in \cite{ASMST}, where $K$=$150$ and $L$=$10$.}
	\label{figure-subpacketization}
\end{figure}

Let us take some numerical comparisons in Table \ref{tab-subpacketization} to verify our theoretical comparisons.  By Table \ref{tab-subpacketization}, the subpacketizations of our schemes in Table \ref{tab-Theorem-2} are smaller than those of the ASMST scheme.
{\begin{table*}[htbp]
		\caption{The subpacketizations of the ASMST scheme in \cite{ASMST} and our schemes in Table \ref{tab-Theorem-2}.}
		\centering
		\renewcommand\arraystretch{1.2}{
			\setlength{\tabcolsep}{1.2mm}{
				\begin{tabular}{|c|c|c|c|c|c|c|}\hline
					\multicolumn{3}{|c|}{ } &\multicolumn{4}{|c|}{subpacketization}\\\cline{4-7}
					\multicolumn{3}{|c|}{} & \multicolumn{1}{|c|}{ASMS scheme} &\multicolumn{3}{|c|}{Table \ref{tab-Theorem-2}} \\
					\hline
					$K$ & $M/N$ & $L$ & $F_{\text{ASMST}}$ & \text{Scheme 1}&  \text{Scheme 2}&\text{Scheme 3} \\
					\hline
					20 & 0.2 & 4 & 2204475 &5&20&20  \\
					\hline
					20 & 0.4 & 5 & 20785050 &10  &30 &20   \\
					\hline
					50 & 0.2 & 5 & 8.4E$+$14 & 45 &360 & 50  \\
					\hline
					50 & 0.3 & 5 & 1.0E$+$17 & 120 &840 & 50  \\
					\hline
					100 & 0.05 & 5 & 2.3E$+$14 & 20 &380 &100   \\
					\hline
					100 & 0.2 & 10 & 1.1E$+$32 &45 &360 & 100   \\
					\hline
					150 & 0.06 & 10 & 4.8E$+$28 &15  &210 & 150   \\
					\hline
					150 & 0.1 & 15 & 5.5E$+$38 & 10 &90 & 150  \\
					\hline
					150 & 0.2 & 15 & 1.9E$+$49 & 45 &360 &150  \\
					\hline
		\end{tabular}}}\label{tab-subpacketization}
\end{table*}} 
 In addition, we propose another comparisons for the case $K=150$, $L=10$ and $m=10$ in Figure \ref{figure-subpacketization}. Compared to the ASMST scheme in \cite{ASMST},  Schemes 1 and Scheme 2 have a very small subpacketization.

Finally let us consider the computational complexity of the ASMST scheme and our schemes. In the ASMST scheme, the computational complexity in the whole communication process is $\lambda_{\text{ASMST}}
=O((t+L) {t+L-1\choose t}^{3}{K\choose t+L} )$. In our schemes, by Lemma \eqref{lemma2-MAPDAs}, the computational complexity in the whole communication process can be represented as 
\begin{itemize}
\item $\lambda_{\text{Scheme 1}}=(O((t+L)^3)+O((t+L)^{2})+O(t(t+L)) )\cdot \frac{K-t}{\alpha}{K/\alpha\choose (t+L)/\alpha} \approx O\left((t+L)^3\cdot \frac{K-t}{\alpha}{K/\alpha\choose (t+L)/\alpha}\right)$;
\item $\lambda_{\text{Scheme 2}}=\left(O((t+L)^3)+O((t+L)^{2})+O(t(t+L)) \right)\cdot \text{sgn}(\frac{t}{m}+1,\frac{m}{L})\cdot l\cdot\frac{K-t}{t+m} {K/m\choose t/m}\approx O\left(t+L)^3\cdot{K/m\choose t/m}\right)$;
\item $\lambda_{\text{Scheme 3}}=(O((t+L)^3)+O((t+L)^{2})+O(t(t+L)) )\cdot (K-t)\approx O\left((t+L)^3\cdot(K-t)\right)$.
\end{itemize}
Then we have the ratio of $\lambda_{\text{ASMST}}$ and $\lambda_{\text{Scheme 1}}$ as follows.
\begin{equation} \label{complexity-ratio-1}
\begin{split}
\frac{\lambda_{\text{ASMST}}}{\lambda_{\text{Scheme 1}}}=&\frac{O\left((t+L) {t+L-1\choose t}^{3}{K\choose t+L}\right)}{O\left((t+L)^3\cdot \frac{K-t}{\alpha}{K/\alpha\choose (t+L)/\alpha}\right)}
\approx O\left( 
\frac{(t+L-1)^{3t}\cdot K^{t+L}}{(t+L)^2\cdot(K/\alpha)^{\frac{t+L}{\alpha}} }\right) 
\ \ \ \ (K\rightarrow \infty)\\
\approx&O((t+L-1)^{3t-2}\cdot \alpha^{\frac{t+L}{\alpha}}\cdot K^{(t+L)\cdot\frac{\alpha-1}{\alpha}} ).
\ \ \ \ (K\rightarrow \infty)
	\end{split}
\end{equation} 
From \eqref{complexity-ratio-1}, the computational complexity of the ASMST scheme is about $O((t+L-1)^{3t-2}\cdot \alpha^{\frac{t+L}{\alpha}}\cdot
K^{(t+L)\cdot\frac{\alpha-1}{\alpha}} )$ times larger than that of Scheme 1 when $K$ is sufficiently large. Similar to \eqref{complexity-ratio-1}, when $K$ is large, the computational complexity of the ASMST scheme is about $O((t+L-1)^{3t-2}\cdot m^{\frac{t}{m}}\cdot K^{L+t\cdot\frac{m-1}{m}} )$ times larger than that of Scheme 2, and the computational complexity of the ASMST scheme is about 
$O((t+L-1)^{3t-2}\cdot K^{L+t-1} )$ times larger than that of Scheme 3. 
Therefore, the computational complexity of our schemes is much lower than that of the ASMST scheme.

\subsection{Sketch of designing the scheme in Theorem \ref{th-1}}
Given a $(L,K,F,Z,S)$ PDA $\mathbf{P}$, we use the $K$ columns and $F$ rows denote the users and packets of each file, respectively. The entry $\mathbf{P}(f,k)=*$ represents that the $f^{\text{th}}$ packet of all files is cached by user $k$. Each user caches $M=\frac{ZN}{F}$ files by Condition C$1$ of Definition \ref{def-MAPDA}. If $\mathbf{P}(f,k)=s$ is an integer, it means that the $f^{\text{th}}$ packet of all files is not stored by user $k$. In the uplink step, each user transmits the linear combination of all the requested packets indicated by $s$ to the BS at transmission $s$; in the downlink step, the BS transmits the received signal directly to the users through the downlink channel. Condition C$3$ of Definition \ref{def-MAPDA} guarantees that each user can receive at most one of its requested packets, and the Condition C$4$ guarantees that each user can recover its required packets contained in a received signal since it can cancel all the other packets with the help of its cached packets. Since there are $F-Z$ required packets, then the sum-DoF is $\frac{K(F-Z)}{S}$. Then we can obtain an $F$-division $(L,K,M,N)$ MIR coded caching scheme with memory size $M=\frac{ZN}{F}$ and NDT  $\tau=\frac{S}{F}$. Recall that we omit the coding process  $\psi$  and the noises in the uplink and the downlink steps.

Now let us take  $(L,K,F,Z,S)$ = $(2,6,3,1,3)$ MAPDA $\mathbf{P}$ in Example \ref{ex-1} to illustrate our generating method for  the $3$-division $(L, K, M, N)$ = $(2,6,2,6)$ MIR coded caching scheme as follows.

\noindent $\bullet$ {\bf Placement phase:} Each file $\mathbf{W}_{n}$ is divided into $3$ packets, i.e., $\mathbf{W}_{n}=\left\lbrace \mathbf{W}_{n,1}, \mathbf{W}_{n,2}, \mathbf{W}_{n,3}\right\rbrace  $. Specifically, each user $k$ caches the packets $\mathbf{W}_{n,f} $ if the entry $\mathbf{P}(f,k)=* $, the caches of $6$ users are
\begin{align}
	\mathcal{Z}_{1}=\mathcal{Z}_{4}=\{\mathbf{W}_{n,1} |  n\in[6]\},\ \ 
	\mathcal{Z}_{2}=\mathcal{Z}_{5}=\{\mathbf{W}_{n,2} |  n\in[6]\},\ \ 
	\mathcal{Z}_{3}=\mathcal{Z}_{6}=\{\mathbf{W}_{n,3} |  n\in[6]\}.
	\label{eq-cache-k}
\end{align}Clearly each use cache the packets of size $M=\frac{1\times N}{F}=2$ files.

\noindent $\bullet$  {\bf Delivery phase:} We assume that request vector ${\bf d}=(1,2,3)$. This phase contains the uplink and downlink steps, each of which contains $3$ transmissions. We first consider the first transmission, i.e.,  $s=1$. In $\mathbf{P}$, we have $\mathbf{P}(2,1)=\mathbf{P}(1,2)=\mathbf{P}(2,4)=\mathbf{P}(1,5)=1$. In the uplink step, the served users and their requested packets are
\begin{equation*}
	\label{eq-packet-user-time-s=1}
	\mathcal{K}_1=\{1,2,4,5\} \ \text{and} \ \mathbf{\mathbf{W}}^{(1)}=\left(
	\begin{array}{c}
		\mathbf{W}_{1,2}\\
		\mathbf{W}_{2,1}\\
		\mathbf{W}_{4,2}\\
		\mathbf{W}_{5,1}
	\end{array}
	\right),
\end{equation*}	respectively. Then user $1$, $2$, $4$ and $5$ send the coded signals $x_1$, $x_2$, $x_4$ and $x_5$ respectively. That is,  
\begin{align*}
	\begin{split}
		\mathbf{X}^{(1)}&=
		\left(
		\begin{array}{c}
			\mathbf{x}_{1}(1)\\
			\mathbf{x}_{2}(1)\\
			\mathbf{x}_{4}(1)\\
			\mathbf{x}_{5}(1)\\
		\end{array}
		\right)=\mathbf{V}^{(1)}\mathbf{W}^{(1)},
		\label{eq:X in transmission-1}
	\end{split}
\end{align*}where the precoding matrix can be obtained by the following channel matrix 

	\begin{align*}
		\mathbf{H}^{(1)}=({\bf h}^{(1)}_1, {\bf h}^{(1)}_2, {\bf h}^{(1)}_4, {\bf h}^{(1)}_5)=	\left(
		\begin{array}{cccc}
			1& 1& 1 & 1  \\
			2 & 3 & 4 & 5  
		\end{array}
		\right).
	\end{align*} Then, the BS can receive the following signals with help of its $2$ antennas.
	\begin{align*}
		\mathbf{Y}_{\text{BS}}(1)\!&=\mathbf{H}^{(1)}\mathbf{V}^{(1)}\mathbf{W}^{(1)}
		=\left(
		\begin{array}{cccc}
			1& 1& 1 & 1  \\
			2 & 3 & 4 & 5  
		\end{array}
		\right)
		\left(
		\begin{array}{cccc}
			0& v^{(1)}_{1,2} & 0 & v^{(1)}_{1,4}  \\
			v^{(1)}_{2,1} & 0 & v^{(1)}_{2,3} & 0  \\
			0& v^{(1)}_{3,2} & 0 & v^{(1)}_{3,4}  \\
			v^{(1)}_{4,1} & 0 & v^{(1)}_{4,3} & 0  
		\end{array}
		\right)
		\left(
		\begin{array}{c}
			\mathbf{W}_{1,2}\\
			\mathbf{W}_{2,1}\\
			\mathbf{W}_{4,2}\\
			\mathbf{W}_{5,1}
		\end{array}
		\right),
	\end{align*} and directly sends to the users in $\mathcal{K}$ in the downlink step. Then these $4$ users can receive the following signals respectively.
	\begin{eqnarray*}
		\begin{split}
			\mathbf{Y}(1)
			&=\left(
			\begin{array}{c}
				\mathbf{y}_1(1)\\
				\mathbf{y}_2(1)\\
				\mathbf{y}_4(1)\\
				\mathbf{y}_5(1)\\
			\end{array}
			\right)=
			\left( \mathbf{H}^{*}\right) ^{(1)}
			\mathbf{H}^{(1)}
			\mathbf{V}^{(1)}
			\left(
			\begin{array}{c}
				\mathbf{W}_{1,2}\\
				\mathbf{W}_{2,1}\\
				\mathbf{W}_{4,2}\\
				\mathbf{W}_{5,1}
			\end{array}
			\right)\\
			&=	\left(
			\begin{array}{ccccc}
				1 & \frac{3}{2} & 0 & -\frac{1}{2}\\
				\frac{1}{2} & 1 & \frac{1}{2} & 0 \\
				0 & \frac{1}{2} & 1 & \frac{1}{2} \\
				-\frac{1}{2} & 0 & \frac{3}{2} & 1 
			\end{array}
			\right)		\left(
			\begin{array}{c}
				\mathbf{W}_{1,2}\\
				\mathbf{W}_{2,1}\\
				\mathbf{W}_{4,2}\\
				\mathbf{W}_{5,1}
			\end{array}
			\right)=	\left(
			\begin{array}{c}
				\mathbf{W}_{1,2}+\frac{3}{2}\mathbf{W}_{2,1} -\frac{1}{2}\mathbf{W}_{5,1}\\
				\frac{1}{2}\mathbf{W}_{1,2}	+\mathbf{W}_{2,1}+\frac{1}{2}\mathbf{W}_{4,2}\\
				\frac{1}{2}\mathbf{W}_{2,1}+\mathbf{W}_{4,2}+\frac{1}{2}\mathbf{W}_{5,1}\\	
				-\frac{1}{2}\mathbf{W}_{1,2}+\frac{3}{2}\mathbf{W}_{4,2}+\mathbf{W}_{5,1}	
			\end{array}
			\right).
			\label{eq-users-received signals}
		\end{split}
	\end{eqnarray*}By 
	\begin{align*}
		\left( \mathbf{H}^{*}\right) ^{(1)}
		\mathbf{H}^{(1)}\mathbf{V}^{(1)}=\left(
		\begin{array}{ccccc}
			1 & \frac{3}{2} & 0 & -\frac{1}{2}\\
			\frac{1}{2} & 1 & \frac{1}{2} & 0 \\
			0 & \frac{1}{2} & 1 & \frac{1}{2} \\
			-\frac{1}{2} & 0 & \frac{3}{2} & 1 
		\end{array}
		\right)	,
	\end{align*} we have the precoding matrix
	\begin{align*}
		\mathbf{V}^{(1)}=\left(
		\begin{array}{cccc}
			0& \frac{21}{4} & 0 & -\frac{13}{4}  \\
			\frac{21}{4} & 0 & -\frac{11}{4} & 0  \\
			0& -\frac{11}{4} & 0 &\frac{7}{4}  \\
			-\frac{13}{4} & 0 & \frac{7}{4} & 0  
		\end{array}
		\right)	.
\end{align*}
We can see that user $1$ receives the coded signal $\mathbf{y}_1(1)=\mathbf{W}_{1,2}+\frac{3}{2}\mathbf{W}_{2,1} -\frac{1}{2}\mathbf{W}_{5,1}$. By its cached packets $\mathbf{W}_{2,1}$ and $\mathbf{W}_{5,1}$, user $1$ can decode its required packet $\mathbf{W}_{1,2}$. Similarly user $2$, $4$ and $5$ can decode their required packets $\mathbf{W}_{2,1}$, $\mathbf{W}_{4,2}$ and $\mathbf{W}_{5,1}$, respectively.  The other two transmissions can be treated similarly.	

From \eqref{eq-comput-NDT} and \eqref{eq-comput-NDT-DL}, we can obtain the NDT  $\tau_{\text{UL}}=\tau_{\text{DL}}=\frac{S}{F}=\frac{3}{3}=1$ and the computational complexity in the whole communication process is 
	\begin{align*}
		\lambda_{\text{New}}=\left((t+L)^3+(t+L)^{2}+t(t+L) \right)\cdot S=\left((2+2)^3+(2+2)^{2}+2(2+2) \right)\cdot 3=264.	
	\end{align*}	
By Lemma \ref{lemma-1} we can obtain the $(L,K,M,N)=(2,6,2,6)$  ASMST coded caching scheme with the NDT $\tau_{\text{UL}}=\tau_{\text{DL}}=\frac{K-t}{t+L}=\frac{6-2}{2+2}=1$ and the computational complexity in the whole communication process 
\begin{align*}
\lambda_{\text{ASMST}}&=\left( {t+L-1\choose t}(t+1)(L-1) +  {t+L\choose t+1}L  + 2L{t+L\choose t+1}+ (t+L) {t+L-1\choose t}^{3}  \right) {K\choose t+L}\\
&=\left( {2+2-1\choose 2}(2+1)(2-1) +  {2+2\choose 2+1}2  + 2\cdot 2{2+2\choose 2+1}+ (2+2) {2+2-1\choose 2}^{3}  \right) {6\choose 2+2}=2115. 
\end{align*}Clearly, the computationnal complexity of our scheme is $\frac{\lambda_{\text{New}}}{\lambda_{\text{ASMST}}}=\frac{264}{2115}=0.1248$ times smaller than that of the ASMST scheme with the same NDT.

\section{The proof of Theorem \ref{th-1}}\label{proof}
Suppose that $\mathbf{P}$ is an $(L, K, F, Z, S)$
MAPDA with $KZ\geq LF$, we can obtain an $F$-division $(L, K, M, N)$ MIR coded caching scheme with memory size $M=\frac{ZN}{F}$ and the optimal $\tau_{\text{UL}}=\tau_{\text{DL}}=\frac{S}{F}$ as follows.
\subsection{The Scheme Realized by $\mathbf{P}$}
$\bullet$ {\bf Placement phase:} Each file $\mathbf{W}_{n} $ is divided into $F$ packets, i.e., $\mathbf{W}_{n}=(\mathbf{W}_{n,f}|f\in [F]
)$. Each user $k$ caches the following content.
\begin{align}
\mathcal{Z}_{k}=\left\{\mathbf{W}_{n,f}\ \Big|\ \mathbf{P}(f,k)=*, n\in [N], f\in[F] \right\}
\end{align}Clearly each user caches exactly $M=\frac{ZN}{F}$ files.

$\bullet$  {\bf Delivery phase:} Assume the demand vector ${\bf d}=(d_1,d_2,\ldots,d_K)$. This phase contains the uplink and downlink steps, each of which contains $S_{\text{UL}}=S_{\text{DL}}=S$ transmissions. 
\begin{itemize}
\item[--]{\bf In the uplink step:} Recall that the maximum sum-DoF of $\mathbf{P}$ is $t+L$, each integer appears exactly $t+L$ times in $\mathbf{P}$. For any $s\in[S]$, assume that $\mathbf{P}(f_{1},k_{1})= \mathbf{P}(f_{2},k_{2})=\ldots=\mathbf{P}(f_{t+L},k_{t+L})=s$, which represents that user $k_i$ requests the packet $\mathbf{W}_{d_{k_i},f_i}$. Thus the served users and their requested packets are
\begin{equation*}\label{eq-packet-user-time-s}
\mathcal{K}_s=\{{k}_{1},\!{k}_{2},\!\ldots,\!{k}_{t+L}\}\ \text{and} \ \mathbf{W}^{(s)}=\left(
\begin{array}{c}
\mathbf{W}_{d_{{k}_{1}},f_{1}}\\
\mathbf{W}_{d_{{k}_{2}},f_{2}}\\
\vdots\\
\mathbf{W}_{d_{{k}_{t+L}},f_{t+L}}
\end{array}
\right),
\end{equation*}
respectively. Each user $k_i \in[K]$ for $i\in [t+L]$ transmits the following linear combination of requested packets
\begin{equation}\label{MAMIR-sig}
\mathbf{x}_{k_{i}}(s)=\sum_{j=1}^{t+L}v_{i,j}^{(s)}\mathbf{W}_{d_{k_{j}},f_j},
\end{equation}
We denote $\mathbf{v}_{j}(s)=\left( v_{1,j}(s),\!\ldots,\!v_{t+L,j}(s)\right)$. The users indices in $\mathcal{K}_s $ who neither cache nor request packet $\mathbf{W}_{d_{k_{i}},f_i} $ are denoted by $\mathcal{C}_{i}(s) $. We design $\mathbf{v}_{i}^{T}(s)$ as the right null vector of $\mathbf{A}\left( \mathcal{C}_{i}(s), [t+L]\right)  $, i.e., $\mathbf{A}\left( \mathcal{C}_{i}(s), [t+L]\right)\mathbf{v}_{i}^{T}(s)=0 $, where $\mathbf{A}$ specifically represents in \eqref{eq-users-receive}. In addition, in each transmission, the computational complexity of computing $ \mathbf{x}_{k_{i}}(s)$ is $O((t+L)^{2}) $. Then, the BS receives the signal
\begin{align}\label{MAMIR-BS}
\mathbf{y}_{\text{BS}}(s)& = \sum\limits_{i=1}^{t+L}\mathbf{h} _{k_{i}} {\bf x}_{k_{i}}(s),
\end{align}

\item[--]{\bf In the downlink step:} In the transmission $s\in [S]$, the BS directly transmits the received signal $\mathbf{y}_{\text{BS}}(s)$ to the users through the downlink channel. Thus, user $k_l$, where $l\in [t+L]$, receives the signal
\begin{eqnarray}
\begin{split}
{\bf y}_{k_l}^{(s)}
=\mathbf{h}^{*}_{k_{l}}\mathbf{y}_{\text{BS}}(s)=\mathbf{h}^{*}_{k_{l}}\sum\limits_{i=1}^{t+L} \mathbf{h}_{k_{i}}\sum_{j=1}^{t+L}v_{i,j}^{(s)}\mathbf{W}_{d_{k_{j}},f_j}=\sum_{j=1}^{t+L} \left(\sum\limits_{i=1}^{t+L} \mathbf{h}^{*}_{k_{l}}\mathbf{h}_{k_{i}}v_{i,j}^{(s)}\right)\mathbf{W}_{d_{k_{j}},f_j}.
\end{split}
\label{eq-receive-user-k_{l}}
\end{eqnarray}

Then from \eqref{MAMIR-sig}, \eqref{MAMIR-BS} and \eqref{eq-receive-user-k_{l}}, the received signals by all $t+L $ users in transmission $s$ are
\begin{eqnarray}
\begin{split}
\mathbf{Y}(s)
&=\left(
\begin{array}{c}
\mathbf{h}_{k_1}^{*}\\
\mathbf{h}_{k_2}^{*}\\
\vdots\\
\mathbf{h}_{k_{t+L}}^{*}
\end{array}
\right)
(\mathbf{h}_{k_{1}},\mathbf{h}_{k_{2}},\ldots,\mathbf{h}_{k_{t+L}})
\left(
\begin{array}{cccc}
v_{1,1}^{(s)}&v_{1,2}^{(s)}&\cdots&v_{1,t+L}^{(s)}\\
v_{2,1}^{(s)}&v_{2,2}^{(s)}&\cdots&v_{2,t+L}^{(s)}\\
\vdots&\vdots&\ddots&\vdots\\
v_{t+L,1}^{(s)}&v_{t+L,2}^{(s)}&\cdots &v_{t+L,t+L}^{(s)}
\end{array}
\right)
\left(
\begin{array}{c}
\mathbf{W}_{d_{{k}_{1}},f_{1}}\\
\mathbf{W}_{d_{{k}_{2}},f_{2}}\\
\vdots\\
\mathbf{W}_{d_{{k}_{t+L}},f_{t+L}}
\end{array}
\right),
\end{split}
\label{eq-users-receive}
\end{eqnarray}
where \begin{align*}
\left(
\begin{array}{c}
\mathbf{h}_{k_1}^{*}\\
\mathbf{h}_{k_2}^{*}\\
\vdots\\
\mathbf{h}_{k_{t+L}}^{*}
\end{array}
\right)
(\mathbf{h}_{k_{1}},\mathbf{h}_{k_{2}},\ldots,\mathbf{h}_{k_{t+L}})=\mathbf{H}^{*}(\mathcal{K}_s, [L])\mathbf{H}([L],\mathcal{K}_s)=\mathbf{A}.
\end{align*}
\end{itemize}

In the proposed scheme, we design the beamforming vector $v_{i,j}^{(s)}$ to enable one-shot delivery, ensuring that each requested packet can be directly decoded by the desired users. The details of this design are presented in the following subsection.

\subsection{Decodability for Each User}
For each $j\in [t+L]$ , we suppose that there are $\lambda^{(s)}_{j}$ rows with indices in $[t+L]$ containing integers at the column $k_{l}$ of $\mathbf{P}$. Therefore, the set of these rows indices can be written as
\begin{eqnarray}
\label{eq-not-cached}
\mathcal{J}^{(s)}_{l}=\left\{j\ |\ \mathbf{P}(f_{j},{k}_{l})\neq*, j\in [t+L]\setminus\left\lbrace l\right\rbrace \right\}.
\end{eqnarray}
Obviously, $|\mathcal{J}^{(s)}_{l}|=\lambda^{(s)}_{j}$.
Thus, \eqref{eq-receive-user-k_{l}} can be represented as
\begin{align}\label{eq-receive-users-k-subset}
{\bf y}_{k_l}^{(s)}&=\sum_{j=1}^{t+L} \left(\sum\limits_{i=1}^{t+L} \mathbf{h}^{*}_{k_{l}}\mathbf{h}_{k_{i}}v_{i,j}^{(s)}\right)\mathbf{W}_{d_{k_{j}},f_j}=\underbrace{\left(\sum\limits_{i=1}^{t+L} \mathbf{h}^{*}_{k_{l}}\mathbf{h}_{k_{i}}v_{i,l}^{(s)}\right)\mathbf{W}_{d_{k_{l}},f_l}}_{\text{Required} \  \& \ \text{Uncaching packet}}
\\
&+\underbrace{\sum\limits_{j\in\mathcal{J}^{(s)}_{l} }
\!\! \left(\sum\limits_{i=1}^{t+L} \mathbf{h}^{*}_{k_{l}}\mathbf{h}_{k_{i}}v_{i,j}^{(s)}\right)\mathbf{W}_{d_{k_{j}},f_j}}_{\text{Unrequired}\  \& \ {\text{Uncaching packets}}} +\underbrace{\sum\limits_{j\in[t+L]\setminus\left( \left\lbrace l\right\rbrace \cup\mathcal{J}^{(s)}_{l}\right) } \left(\sum\limits_{i=1}^{t+L} \mathbf{h}^{*}_{k_{l}}\mathbf{h}_{k_{i}}v_{i,j}^{(s)}\right)\mathbf{W}_{d_{k_{j}},f_j}}
_{\text{Caching packets}},\nonumber
\end{align}
where the packet in the first term is requested by user $k_{l}$; the packets in the second term are neither required nor cached by user $k_{l}$; the packets in the third term are not required but cached by user $k_{l}$. Obviously, only the packets in the first two terms in \eqref{eq-receive-users-k-subset} need to be considered, as user $k_{l}$ can cancel all the packets in the third term with its cached contents. To decode the desired packet $\mathbf{W}_{d_{k_{l}},f_{l}}$, it is necessary to cancel the interfering packets in the second term of \eqref{eq-receive-users-k-subset}. In particular, \eqref{eq-users-receive} can also be written as follws.
\begin{eqnarray}
\label{Y(s)}
\begin{split}
\mathbf{Y}(s)&=\left( \mathbf{H}^{*}\right) ^{(s)}\mathbf{H}^{(s)}\mathbf{X}(s)
=\left( \mathbf{H}^{*}\right) ^{(s)}\mathbf{H}^{(s)}\mathbf{V}^{(s)}\mathbf{W}^{(s)} \\
&=\left( \mathbf{H}^{*}\right) ^{(s)}\mathbf{H}^{(s)}\left({\bf v}^{(s)}_{1},{\bf v}^{(s)}_{2},\ldots,{\bf v}^{(s)}_{t+L}\right)
\mathbf{W}^{(s)} \\
&=\left(\left( \mathbf{H}^{*}\right) ^{(s)}\mathbf{H}^{(s)}{\bf v}^{(s)}_{1},\left( \mathbf{H}^{*}\right) ^{(s)}\mathbf{H}^{(s)}{\bf v}^{(s)}_{2},\ldots,\left( \mathbf{H}^{*}\right) ^{(s)}\mathbf{H}^{(s)}{\bf v}^{(s)}_{t+L} \right)
\mathbf{W}^{(s)} \\
&=\mathbf{B}^{(s)}\mathbf{W}^{(s)}.
\end{split}
\end{eqnarray}
Note that $ \mathbf{H}^{*} $ denotes the conjugate-transpose of $\mathbf{H}$, any sub matrix of $ \mathbf{H}$ is full rank. Clearly, user $k_{l}$ can decode its required packet $\mathbf{W}_{d_{k_{l}},f_{l}}$ by its received signal ${\bf y}_{k_l}^{(s)}$ and its cached packets $\mathcal{Z}_{k_l}$ if the following condition holds for any two different integers $l,j\in [t+L]$.
\begin{align}
\label{eq-coding-vector2}
\left\{\begin{array}{c}
1=\sum\limits_{i=1}^{t+L} \mathbf{h}^{*}_{k_{l}}\mathbf{h}_{k_{i}}v_{i,l}^{(s)}
=\mathbf{h}^{*}_{k_{l}}\mathbf{H}^{(s)}{\bf v}^{(s)}_{l}=
\mathbf{B}^{(s)}(l,l) ,\ \ \ \ \ \ \ \ \ \ \ \ 
\\
0=\sum\limits_{i=1}^{t+L} \mathbf{h}^{*}_{k_{l}}\mathbf{h}_{k_{i}}v_{i,j}^{(s)}=\mathbf{h}^{*}_{k_{l}}\mathbf{H}^{(s)}{\bf v}^{(s)}_{j}=
\mathbf{B}^{(s)}(l,j), j\in \mathcal{J}^{(s)}_{l}.
\end{array}\right.
\end{align}
Note that the first equality in \eqref{eq-coding-vector2} means that user $k_l$ can decode its desired  packet $\mathbf{W}_{d_{k_{l}},f_{l}}$, while the second equality in \eqref{eq-coding-vector2} means that the user $k_l$ is able to cancel its un-required and un-cached coded packet $\mathbf{W}_{d_{k_{j}},f_{j}}$.
Clearly, in the entire communication process, each user can decode all packets of its required file if \eqref{eq-coding-vector2} holds for all $s\in [S]$. So, it is sufficient to show that there exists a precoding matrix $\mathbf{V}^{(s)}$ that satisfies \eqref{eq-coding-vector2}.

In each transmission, $t+L$ users can decode their requested packets, the computational complexity is $O\left(t(t+L)\right) $. 
The design of the precoding matrices, which is equivalent to computing $\mathbf{V}^{(s)}=(( \mathbf{H}^{*}) ^{(s)}\mathbf{H}^{(s)})^{-1}\mathbf{B}^{(s)}$. Thus the computational complexity of computing the precoding matrix for each transmission is $O((t+L)^3)$. From \eqref{Y(s)} and \eqref{eq-coding-vector2}, the singles received by $t+L $ users can also be expressed as 
\begin{eqnarray}
\begin{split}
\mathbf{Y}(s)&=\mathbf{B}^{(s)}\mathbf{W}^{(s)}:=\left(
\begin{array}{cccc}
b_{1,1}^{(s)}&b_{1,2}^{(s)}&\cdots&b_{1,t+L}^{(s)}\\
b_{2,1}^{(s)}&b_{2,2}^{(s)}&\cdots&b_{2,t+L}^{(s)}\\
\vdots &\vdots&\ddots&\vdots\\
b_{t+L,1}^{(s)}&b_{t+L,2}^{(s)}&\cdots&b_{t+L,t+L}^{(s)}\\
\end{array}
\right)
\left(
\begin{array}{c}
\mathbf{W}_{d_{{k}_{1}},f_{1}}\\
\mathbf{W}_{d_{{k}_{2}},f_{2}}\\
\vdots\\
\mathbf{W}_{d_{{k}_{t+L}},f_{t+L}}
\end{array}
\right),
\end{split}
\label{eq-users-received}
\end{eqnarray}
where \begin{eqnarray*}
b_{m,n}^{(s)}=\left\{
\begin{array}{cc}
1 &\text{if}\ m=n  \\
0 &\text{if}\ m\neq n, \mathbf{P}(f_{n},k_{m})\neq *.
\end{array}
\right.
\end{eqnarray*}

\subsection{The Existence of $\mathbf{V}^{(s)}$ that Satisfies \eqref{eq-coding-vector2}}
Next, we examine the existence of a matrix $\mathbf{V}^{(s)}$ that satisfies the equation \eqref{eq-coding-vector2}. To this end, we select the appropriate linear coding coefficients $v_{i,j}$ for all $i \in [t+L]$ and $j \in [t+L]$ such that \eqref{eq-coding-vector2} holds consistently.
From \eqref{T-sig}, if user $k_i$ does not cache the packet $\mathbf{W}_{d_{k_{j}},f_j}$, then $v_{i,j}^{(s)}=0$. Besides, for the packet $\mathbf{W}_{d_{k_{j}},f_j}$, according to $\mathbf{P}^{(s)}$, we know that $t$ is in $ t+L$ users cache it, so the number of complex numbers in each column of the precoding matrix $\mathbf{V}^{(s)}$, i.e., the number of unknowns, is $t$; for the packet $W_{d_{k_{j}},f_j}$, according to $\mathbf{P}^{(s)}$, we can know that $L$ is in $ t+L$ users do not cache it, so in the receiving matrix $\mathbf{B}^{(s)} $, the diagonal elements are all $1$'s, and the number of $1$'s and 0's in each column, i.e., the number of equations, is $L $. The necessary and sufficient condition for $\mathbf{V}^{(s)}$ to exist is that the equation \eqref{equation-solution} has a solution. In this case, the sufficient condition for the equation \eqref{equation-solution} to have a solution is $t \geq L$.

\begin{eqnarray}\label{equation-solution}
\begin{split}
\left(
\begin{array}{c}
\mathbf{h}_{k_1}^{*}\\
\mathbf{h}_{k_2}^{*}\\
\vdots\\
\mathbf{h}_{k_{t+L}}^{*}
\end{array}
\right)
(\mathbf{h}_{k_{1}},\mathbf{h}_{k_{2}},\ldots,\mathbf{h}_{k_{t+L}})
\left(
\begin{array}{cccc}
v_{1,1}^{(s)}&v_{1,2}^{(s)}&\cdots&v_{1,t+L}^{(s)}\\
v_{2,1}^{(s)}&v_{2,2}^{(s)}&\cdots&v_{2,t+L}^{(s)}\\
\vdots&\vdots&\ddots&\vdots\\
v_{t+L,1}^{(s)}&v_{t+L,2}^{(s)}&\cdots &v_{t+L,t+L}^{(s)}
\end{array}
\right)
\left(
\begin{array}{c}
\mathbf{W}_{d_{{k}_{1}},f_{1}}\\
\mathbf{W}_{d_{{k}_{2}},f_{2}}\\
\vdots\\
\mathbf{W}_{d_{{k}_{t+L}},f_{t+L}}
\end{array}
\right)
=\\
\left(
\begin{array}{cccc}
b_{1,1}^{(s)}&b_{1,2}^{(s)}&\cdots&b_{1,t+L}^{(s)}\\
b_{2,1}^{(s)}&b_{2,2}^{(s)}&\cdots&b_{2,t+L}^{(s)}\\
\vdots &\vdots&\ddots&\vdots\\
b_{t+L,1}^{(s)}&b_{t+L,2}^{(s)}&\cdots&b_{t+L,t+L}^{(s)}\\
\end{array}
\right)
\left(
\begin{array}{c}
\mathbf{W}_{d_{{k}_{1}},f_{1}}\\
\mathbf{W}_{d_{{k}_{2}},f_{2}}\\
\vdots\\
\mathbf{W}_{d_{{k}_{t+L}},f_{t+L}}
\end{array}
\right).
\end{split}
\end{eqnarray}

\subsection{Proof of The Existence of $\mathbf{V}^{(s)}$ }
First, proving that  the equation \eqref{equation-solution} has a solution is equivalent to proving that the equation \eqref{equation-solution1} has a solution, and this can further be shown to be equivalent to proving that the equation \eqref{equation-solution2} has a solution, for any $n\in [t+L]$.
\begin{eqnarray}\label{equation-solution1}
\begin{split}
\left(
\begin{array}{c}
\mathbf{h}_{k_1}^{*}\\
\mathbf{h}_{k_2}^{*}\\
\vdots\\
\mathbf{h}_{k_{t+L}}^{*}
\end{array}
\right)
(\mathbf{h}_{k_{1}},\mathbf{h}_{k_{2}},\ldots,\mathbf{h}_{k_{t+L}})
\left(
\begin{array}{cccc}
v_{1,1}^{(s)}&v_{1,2}^{(s)}&\cdots&v_{1,t+L}^{(s)}\\
v_{2,1}^{(s)}&v_{2,2}^{(s)}&\cdots&v_{2,t+L}^{(s)}\\
\vdots&\vdots&\ddots&\vdots\\
v_{t+L,1}^{(s)}&v_{t+L,2}^{(s)}&\cdots &v_{t+L,t+L}^{(s)}
\end{array}
\right)
=
\left(
\begin{array}{cccc}
b_{1,1}^{(s)}&b_{1,2}^{(s)}&\cdots&b_{1,t+L}^{(s)}\\
b_{2,1}^{(s)}&b_{2,2}^{(s)}&\cdots&b_{2,t+L}^{(s)}\\
\vdots &\vdots&\ddots&\vdots\\
b_{t+L,1}^{(s)}&b_{t+L,2}^{(s)}&\cdots&b_{t+L,t+L}^{(s)}\\
\end{array}
\right),
\end{split}
\end{eqnarray}

 \begin{eqnarray}\label{equation-solution2}
\begin{split}
\left(
\begin{array}{c}
\mathbf{h}_{k_1}^{*}\\
\mathbf{h}_{k_2}^{*}\\
\vdots\\
\mathbf{h}_{k_{t+L}}^{*}
\end{array}
\right)
(\mathbf{h}_{k_{1}},\mathbf{h}_{k_{2}},\ldots,\mathbf{h}_{k_{t+L}})
\left(
\begin{array}{cccc}
v_{1,n}^{(s)}\\
v_{2,n}^{(s)}\\
\vdots\\
v_{t+L,n}^{(s)}
\end{array}
\right)=
\left(
\begin{array}{cccc}
b_{1,n}^{(s)}\\
b_{2,n}^{(s)}\\
\vdots \\
b_{t+L,n}^{(s)}\\
\end{array}
\right).
\end{split}
\end{eqnarray}

From MAPDA $\mathbf{P}^{(s)} $, for the packet $\mathbf{W}_{d_{{k}_{n}},f_{n}} $, there are $L$ users among $t+L$ users who not cache the packet. Consequently, there must be $L$ elements in $(v_{1,n}^{(s)}, v_{2,n}^{(s)}, \cdots, v_{t+L,n}^{(s)}) $ that are equal to 0. The  remaining $t$ elements in $(  v_{1,n}^{(s)}, v_{2,n}^{(s)}, \cdots, v_{t+L,n}^{(s)}) $ are assumed to be $( v_{i_{1},n}^{(s)}, v_{i_{2},n}^{(s)}, \cdots, v_{i_{t},n}^{(s)})$. Since user $k_{n}$ requires the packet $\mathbf{W}_{d_{{k}_{n}},f_{n}} $, we have $b_{n,n}^{(s)}=1 $. Excluding user $k_{n}$, $L-1$ users do not cache the packet $W_{d_{{k}_{n}},f_{n}} $, so there are $L-1$ elements in $(b_{1,n}^{(s)}, b_{2,n}^{(s)}, \cdots, b_{t+L,n}^{(s)})$ that are equal to 0. These elements are assumed to be $(b_{l_{1},n}^{(s)}, b_{l_{2},n}^{(s)}, \cdots, b_{l_{L-1},n}^{(s)})=0$.
Therefore, the equation \eqref{equation-solution2} is equivalent to the following equation 
\begin{eqnarray}\label{equation-solution3}
\begin{split}
\left(
\begin{array}{c}
\mathbf{h}_{k_{l_{1}}}^{*}\\
\mathbf{h}_{k_{l_{2}}}^{*}\\
\vdots\\
\mathbf{h}_{k_{l_{L-1}}}^{*}\\
\mathbf{h}_{k_{n}}^{*}
\end{array}
\right)
(
\mathbf{h}_{k_{i_{1}}},\mathbf{h}_{k_{i_{2}}},\ldots,\mathbf{h}_{k_{i_{t}}})
\left(
\begin{array}{cccc}
v_{i_{1},n}^{(s)}\\
v_{i_{2},n}^{(s)}\\
\vdots\\
v_{i_{t},n}^{(s)}
\end{array}
\right)=
\left(
\begin{array}{cccc}
0\\
0\\
\vdots \\
0\\
1
\end{array}
\right),
\end{split}
\end{eqnarray}
In this case, the equation \eqref{equation-solution3} can be expressed as $\mathbf{A}\mathbf{x}=\mathbf{b}  $, where $$\mathbf{A}=\left(
\begin{array}{c}
\mathbf{h}_{k_{l_{1}}}^{*}\\
\mathbf{h}_{k_{l_{2}}}^{*}\\
\vdots\\
\mathbf{h}_{k_{l_{L-1}}}^{*}\\
\mathbf{h}_{k_{n}}^{*}
\end{array}
\right)(
\mathbf{h}_{k_{i_{1}}},\mathbf{h}_{k_{i_{2}}},\ldots,\mathbf{h}_{k_{i_{t}}}).$$
Since any L-order sub-square in 
$\mathbf{H}=(
\mathbf{h}_{k_{1}}~\mathbf{h}_{k_{2}}~\ldots~\mathbf{h}_{k_{t+L}})$ is invertible, 
$$\left(\begin{array}{c}
\mathbf{h}_{k_{l_{1}}}^{*}\\
\mathbf{h}_{k_{l_{2}}}^{*}\\
\vdots\\
\mathbf{h}_{k_{l_{L-1}}}^{*}\\
\mathbf{h}_{k_{n}}^{*}
\end{array}
\right)$$
is invertible. Moreover, when $t\geqslant L $, the $(\mathbf{h}_{k_{i_{1}}},\mathbf{h}_{k_{i_{2}}},\ldots,\mathbf{h}_{k_{i_{t}}})$ 
has full row rank matrix. Thus, we deduce that $\mathbf{A}$ is full-row rank matrix. Since A is a full-row rank, for the equation $\mathbf{A}\mathbf{x}=\mathbf{b}$, we have rank$(\mathbf{A})=rank(\mathbf{A}\lvert \mathbf{b})$. So, we deduce that the equation \eqref{equation-solution3} has a solution, i.e., the equation \eqref{equation-solution} has a solution.

\section{Conclusion}\label{conclusion}
In this paper, we studied the coded caching problem  for the $(K,L,M,N)$ MIR system. Firstly, we found out it is sufficient to focus only on the precoding matrix for the uplink step when $t\geq L$. In this case, we showed that the MAPDA can be used to generate the precoding matrix in the uplink step.
Consequently, based on the existing MAPDAs, we obtained some new schemes for the MIR system which have smaller subpacketization and lower computational complexity compared to the ASMST scheme. We can also obtain the scheme for the case $L>t$ by silencing 
 $L-t$ antennas based on our new schemes.  
\bibliographystyle{IEEEtran}
\bibliography{REFERENCES}

\end{document}